\documentstyle[aps,preprint,amssymb,graphicx]{revtex}
\tightenlines

\begin{document}
\draft \preprint{Accepted for Phys. Rev. A}

\title{Quantum fluid-dynamics from density functional theory}

\author{S.\ K\"ummel$^{1,2}$ and M.\ Brack$^1$}
\address{$^1$Institute for Theoretical Physics, University of
  Regensburg, D-93040 Regensburg, Germany} 
\address{$^2$Department of Physics and Quantum Theory Group, Tulane
  University, New Orleans, Louisiana 70118, USA; e-mail:
skuemmel@tulane.edu}

\date{\today} \maketitle

\begin{abstract}
A partial differential eigenvalue equation for the density
displacement fields associated with electronic excitations is derived
in the framework of density functional theory. Our quantum
fluid-dynamical approach is based on a variational principle and the
Kohn-Sham ground-state energy functional, using only the occupied
Kohn-Sham orbitals. It allows for an intuitive
interpretation of electronic excitations in terms of intrinsic local
currents that obey a continuity equation. We demonstrate the
capabilities of this non-empirical approach by calculating the photoabsorption
spectra of small sodium clusters. The quantitative agreement between
theoretical and experimental spectra shows that even for the smallest
clusters, the resonances observed experimentally at low temperatures
can be interpreted in terms of density vibrations.

\end{abstract}


\narrowtext \flushbottom

\section{Introduction}

Since its formal foundation as a theory of ground-state properties
\cite{hkks}, density functional theory has developed into one of the
most successful methods of modern many-body theory, today also with
well-established extensions such as, e.g., time-dependent \cite{tddft2}
and current \cite{deb,vignale} density-functional theory (DFT). In
particular in the field of metal cluster physics, DFT calculations have
contributed substantially to a qualitative and quantitative understanding
of both ground and excited state properties \cite{overview,chelikowsky}.
Understanding the properties of small metal particles in turn offers
technological opportunities, e.g., to better control catalysis
\cite{hannupalad}, as well as fundamental insights into how matter grows
\cite{akola00,bigpp}. Since the electronic and geometric structure of
metal particles consisting of only a few atoms still cannot be measured
directly, photoabsorption spectra are their most accurate probes.
Especially the spectra of charged sodium clusters have been measured with
high accuracy for a broad range of cluster sizes and temperatures
\cite{divhaberland}. A distinct feature of these spectra is that at
elevated temperatures of several hundred K, in particular for the larger
clusters, only a few broad peaks are observed, whereas at lower temperatures
(100 K and less), a greater number of sharp lines can be resolved for
clusters with only a few atoms. The peaks observed in the high-temperature
experiments found an early and intuitive explanation as collective
excitations in analogy to the bulk plasmon and the giant resonances in nuclei:
different peaks in the spectrum were understood as
belonging to the different spatial directions of the collective motion of the
valence electrons with respect to the inert ionic background. On the other
hand, the sharp lines observed in the low-temperature experiments were       
interpreted as a hallmark of the molecule-like properties of the small
clusters explicable, in the language of quantum chemistry \cite{koutecky96},
only in terms of transitions between molecular states.
 
In this work we present a density functional approach to the calculation
of excitations that leads us to a unified and transparent physical
understanding of the photoabsorption spectra of sodium clusters.
We first derive a general variational principle for the exact energy spectrum
of an interacting many-body system. From this, we derive an approximate
solution in the form of quantum fluid-dynamical differential equations for 
the density displacement fields associated with the electronic vibrations 
around their ground state. By solving these equations, we obtain the
eigenmodes within the DFT; hereby only the ground-state energy 
functional and the occupied Kohn-Sham orbitals are required. 
We demonstrate the accuracy of our
approach by calculating the photoabsorption spectra of small sodium
clusters and comparing our results to low-temperature experiments and
to configuration-interaction (CI) calculations. In this way we can
show that also the spectra of the smallest clusters can be understood,
without knowledge of the molecular many-body wavefunction, in an
intuitive picture of oscillations of the valence-electron density
against the ionic background.
 
\section{A variational principle}

Starting point for the derivation of the variational principle is the
well-known fact that for a many-body system described by a Hamiltonian 
$H$ with ground state $|0\rangle$ and energy $E_0$, the creation
and annihilation operators of all the eigenstates obey the so-called
equations of motion for excitation operators \cite{rowe}
\begin{eqnarray}
\label{equofmotion1}
\langle 0 | {\cal O_\nu}[H , {\cal O}_\nu^\dagger ] | 0 \rangle &=&
\hbar \omega_\nu \langle 0 | {\cal O_\nu O_\nu^\dagger} | 0 \rangle \\ 
\langle 0 | {\cal O_\nu}[H , {\cal O_\nu} ] | 0 \rangle &=& \hbar
\omega_\nu \langle 0 | {\cal O_\nu O_\nu} | 0 \rangle = 0,
\label{equofmotion2} 
\end{eqnarray} 
where ${\cal O_\nu}$ and $\cal{O}_\nu^\dagger$ are defined by
\begin{equation}
  {\cal O}_\nu^\dagger |0\rangle=|\nu\rangle, 
  \hspace{0.5cm}{\cal O_\nu} |\nu\rangle=|0\rangle,
  \hspace{0.5cm}\mbox{and}
  \hspace{0.5cm} {\cal O_\nu}|0\rangle=0.
\end{equation}
Of course, the exact solution of these equations are in general unknown. 
But a variety of approximations to the true excited
states can be derived from them, e.g., the Tam-Dancoff scheme and the
small amplitude limit of time-dependent Hartree-Fock theory (RPA). As
discussed in \cite{rowe}, also higher-order approximations can be obtained. 

Related to these equations, we derive the following
variational principle: solving the equations (\ref{equofmotion1}) and
(\ref{equofmotion2}) for the lowest excited state can be achieved by
solving the variational equation 
\begin{equation}
\label{variation}
\frac{\delta E_3[Q]}{\delta Q}=0
\end{equation}
in the space of all Hermitean operators. Here, $E_3$ is defined by
\begin{equation}
\label{e3}
E_3[Q]=\sqrt{\frac{m_3[Q]}{m_1[Q]}},
\end{equation}
$m_1$ and $m_3$ are the multiple commutators
\begin{eqnarray}
\label{commutator1}
m_1[Q]&=&\frac{1}{2}\langle 0 |\left[Q,\left[H,Q\right]\right]|0
\rangle 
\\ 
m_3[Q]&=&\frac{1}{2}\langle 0
|\left[\left[H,Q\right],\left[\left[H,Q\right],H \right]\right] |0
\rangle,
\label{commutator3}
\end{eqnarray}
and $Q$ is some general Hermitean operator that, as will be shown in
the course of the argument [see Eq.\ (\ref{qpropto}) below], takes the
interpretation of a generalized coordinate. The minimum energy $E_3$
after the variation gives the first excitation energy $\hbar \omega_1$.
The second excitation with energy $\hbar \omega_2$ can be obtained
from variation in an operator space which has been orthogonalized to
the minimum $Q$, and in this way the whole spectrum $\hbar \omega_\nu$ can be
calculated. 
 
The variation $\delta Q$ of an operator can be understood as a
variation of the matrix elements of the operator in the matrix mechanics
picture. Therefore, 
\begin{equation}
  0=\frac{\delta}{\delta
    Q}\left(\frac{m_3[Q]}{m_1[Q]}\right)^\frac{1}{2}=
  \frac{1}{2}\left(\frac{m_3[Q]}{m_1[Q]}\right)^{-\frac{1}{2}}
  \frac{\delta}{\delta Q}\left(\frac{m_3[Q]}{m_1[Q]}\right),
\end{equation} 
and noting that the first factors in the expression to the right are
just $1/(2 E_3)$, 
\begin{equation}
\label{inter1}
0=\frac{\delta}{\delta Q}\left(\frac{m_3[Q]}{m_1[Q]}\right)=
\frac{1}{m_1[Q]}\frac{\delta m_3[Q]}{\delta Q}-\frac{m_3[Q]}{{m_1[Q]}^2}
\frac{\delta m_1[Q]}{\delta Q}
\end{equation} 
is obtained.
With the definition $E_3=\hbar \omega_1$, Eq.\ (\ref{inter1}) turns
into
\begin{equation}
\label{variation2}
  \frac{\delta m_3[Q]}{\delta Q}-(\hbar \omega_1)^2 \frac{\delta
    m_1[Q]}{\delta Q} = 0.
\end{equation}
The variations
\begin{eqnarray}
  \delta m_3[Q]&=&m_3[Q+\delta Q]-m_3[Q] \nonumber \\ \delta
  m_1[Q]&=&m_1[Q+\delta Q]-m_1[Q]
\end{eqnarray}
are evaluated by straightforward application of the commutation rules
(\ref{commutator1}) and (\ref{commutator3}), leading to
\begin{equation}
\label{bigexpect}
\langle 0 | [ \, [\delta Q , H ], \left( [H,[H,Q]] - \left(\hbar
    \omega_1\right)^2 Q \right) \, ] | 0 \rangle = 0.
\end{equation}
With $\delta Q$ Hermitean, $[\delta Q, H]$ is anti-Hermitean, and
(\ref{bigexpect}) therefore is an equation of the form $ c+c^*=0 $
with
\begin{equation}
\label{lastexpect}
c=\langle 0 | [\delta Q , H ] \, \left( [H,[H,Q]] - \left(\hbar
    \omega_1\right)^2 Q \right) | 0 \rangle \in {\Bbb C}.
\end{equation}
Since $| 0 \rangle$ by definition is the exact ground state of $H$,
and (\ref{lastexpect}) must hold for any $\delta Q$, the equation
\begin{equation}
\label{noexpect}
\left( \left[H,\left[H,Q\right]\right] - \left(\hbar \omega_1\right)^2
  Q \right) | 0 \rangle=0
\end{equation} 
is obtained. It resembles the equation of motion for a harmonic oscillator.
Therefore, $Q$ is interpreted as a generalized coordinate, and in analogy to
the well-known algebraic way of solving the harmonic oscillator
problem, $Q$ is written as a linear combination
\begin{equation}
\label{qpropto}
Q \, \propto \, {\cal O}_1^\dagger + {\cal O}_1
\end{equation} 
of the creation and annihilation operator for the first excited state.
Inserting (\ref{qpropto}) into (\ref{noexpect}) leads to the two equations
\begin{eqnarray}
\label{create}
[ H, [ H, {\cal O}_1^\dagger]] | 0 \rangle &=& (\hbar \omega_1)^2{\cal
  O}_1^\dagger | 0 \rangle \\ \left[H,\left[H, {\cal
      O}_1\right]\right] | 0 \rangle &=& (\hbar \omega_1)^2 {\cal O}_1
| 0 \rangle=0.
\label{annihil}
\end{eqnarray} 
First consider (\ref{create}). After closing with state
$\langle 1 |$, one exploits that, by definition, $|0\rangle$ and
$|1\rangle$ are eigenstates of $H$ and evaluates the outer commutator
by letting $H$ act once to the left and once to the right.
Recalling that $\langle 1 | = \langle 0 | {\cal O}_1$, one finally obtains
\begin{equation}
  \langle 0 | {\cal O}_1[H , {\cal O}_1^\dagger ] | 0 \rangle = \hbar
  \omega_1 \langle 0 | {\cal O}_1{\cal O}_1^\dagger | 0 \rangle.
\end{equation} 
This is exactly equation (\ref{equofmotion1}) for the first excited
state. In the same way, (\ref{equofmotion2}) is obtained from
(\ref{annihil}), which completes the derivation of the variational
principle. 

We would like to point out that in earlier work \cite{lrpa}, the RPA
equations have been derived with a related technique that made use of both
generalized coordinate and momentum operators. The advantage of our present 
derivation is that -- although within linear response theory -- it goes
beyond RPA and, due to the formulation in terms of a generalized
coordinate only, is particularly suitable for the formulation of the
variational principle in the framework of density functional theory as 
shown below.

\section{Quantum fluid dynamics from the ground-state energy
functional: a local current approximation}
 
In principle, the exact eigenenergies are defined, via Eqs.\
(\ref{equofmotion1}), (\ref{equofmotion2}), by the variational equation
(\ref{variation}), provided that the operator $Q$ is chosen in a
sufficiently general form. However, just as in the equations of motion
technique, one is forced to make some explicit ansatz for the form of
$Q$, which necessarily introduces approximations.  In Ref.\
\cite{lrpa} it was shown that if $Q$ is taken to be a
one-particle-one-hole excitation operator, Eq.\ (\ref{variation})
leads to the RPA equations. Simplifications of the RPA, in which $Q$
was chosen from restricted sets of local operators $Q_n({\mathbf r})$,
were proposed in connection with both semiclassical \cite{mb89} and
Kohn-Sham density functionals \cite{lrpa}. In the present paper, we
derive a set of quantum fluid-dynamical equations from the variational
principle (\ref{variation}) by choosing $Q$ to a general local
operator $Q({\mathbf r})$. These equations are then solved without any
restriction other than Eq.\ (\ref{defu}) below.

First we recall a relation that is well known in nuclear physics
\cite{bohigas}:
the commutator of Eq.\ (\ref{commutator3}) can be exactly obtained from
\begin{eqnarray}
\label{C}
m_3[Q]
&=&\frac{1}{2}\frac{\partial^2}{\partial
  \alpha^2}\langle\alpha|H|\alpha\rangle \Big|_{\alpha=0},
\end{eqnarray}
where $|\alpha\rangle$ is the state that results from the unitary
transformation 
\begin{equation}
\label{scaltrafo}
|\alpha \rangle = e^{-\alpha S } |0 \rangle, 
\end{equation} 
with $\alpha$ being a real and possibly time-dependent parameter, and
$S$ the so called scaling operator defined by
\begin{equation}
\label{defs2}
S=[H,Q].
\end{equation}
Assuming that $Q$ is just a function of $\mathbf r$ and that the
potentials in $H$ do not contain derivatives with respect to $\bf r$, as is
the case for Coulombic systems, Eq.\ (\ref{defs2}) is easily evaluated:
\begin{equation}
\label{defs}
  S=\sum_{i=1}^{N_{\mathrm e}} s({\bf r}_i)=\sum_{i=1}^{N_{\mathrm e}}
  \frac{1}{2}\left(\nabla_i {\bf u(r}_i)\right) + {\bf u(r}_i)\cdot
  \nabla_i.
\end{equation} 
Here, the displacement field
\begin{equation}
\label{defu}
  {\bf u(r)}=-\frac{\hbar^2}{m}\nabla Q(\bf r)
\end{equation}
has been introduced, and $N_{\mathrm e}$ is the number of electrons.

These equations can be related to DFT by noting that, first, we can 
introduce a set of single particle orbitals
$\{\psi_\mu({\mathbf r}_i)\}$, and from the scaled single particle
orbitals, a scaled single particle density can be constructed via
\begin{equation}
\label{scalen}
n({\mathbf r},\alpha)= \sum_{\mu=1}^{N_{\mathrm e}}
 \left| e^{-\alpha s\left({\bf r}\right)} 
 \psi_\mu({\bf r})\right|^2  
=e^{-\alpha S_n} n({\mathbf r}),
\end{equation}
with a density scaling operator
\begin{equation}
\label{defsn}
S_n=\Big( \nabla \mathbf{u}({\mathbf r})\Big)+
\mathbf{u}({\mathbf r})\cdot\nabla.
\end{equation}
Second, Eq.\ (\ref{commutator1}) can straightforwardly be evaluated
for a local $Q(\mathbf{r})$,
\begin{equation}
\label{m1u}
m_1[Q]=\frac{m}{2\hbar^2} \int {\bf u(r)}\cdot{\bf u(r)}n({\bf r}) 
\, {\mathrm d}^3 r,
\end{equation} 
showing that $m_1$ depends only on $n$ and $\bf u$ and is similar to a
fluid-dynamical inertial parameter. 
And third, we replace the expectation value in Eq.\ (\ref{C}) by
\begin{equation}
\label{todft}
m_3[Q]=\frac{1}{2}\frac{\partial^2}{\partial
  \alpha^2}\langle\alpha|H|\alpha\rangle \Big|_{\alpha=0}
\rightarrow
\frac{1}{2}\frac{\partial^2}{\partial \alpha^2} 
E[n({\bf r}, \alpha)] \Big|_{\alpha=0},
\end{equation}
where $E[n]$ is the usual ground-state Kohn-Sham energy functional
\begin{equation}
\label{clfunctional}
  E[n; \{{\mathbf R}\}]=T_{\mathrm s}[n]+E_{\mathrm xc}[n] +
  \frac{e^2}{2}\int \int
  \frac{n{{\bf(r)}n({\bf r'})}}{\left|{\bf r-r'}\right|} \,\mathrm{d}^3 r' \,
  \mathrm{d}^3 r \nonumber \\ +\int n({\bf r})V_{\mathrm ion}({\bf r;
  \{R\}}) \,\mathrm{d}^3 r.
\end{equation}

Eq.\ (\ref{m1u}) is exact and also Eq.\ (\ref{scalen}) can be verified
order by order, but Eq.\ (\ref{todft}) 
goes beyond the safe grounds on which the energy functional is
defined. However, the replacement of an energy expectation value by the energy
functional is intuitively very plausible, and its practical validity can be
judged {\it a posteriori} by the results. A further strong argument for why
really the density should be the basic variable can be made by calculating the
derivative with respect to time of the scaled density, using Eqs.\
(\ref{scalen}) and (\ref{defsn}), 
\begin{equation}
\label{seecon}
\frac{\mathrm{d}}{\mathrm{d} t}n({\mathbf r},\alpha(t))
=
-S_n \dot{\alpha}(t) \, n({\mathbf r},\alpha(t))
=
-\nabla[ \dot{\alpha}(t)\, {\mathbf u}({\mathbf r}) \,
  n({\mathbf r},\alpha(t))],
\end{equation}
where for the sake of clarity we now explicitly wrote the time dependence of
$\alpha$. Since
\begin{equation}
\label{defj}
  {\bf j}({\bf r}, t)=\dot{\alpha}(t)\, {\mathbf u}({\bf r}) \,
  n({\bf r},\alpha(t)),
\end{equation} 
is a current density, Eq.\ (\ref{seecon}) is just the continuity equation
$
\mathrm{d} n({\bf r},\alpha(t)) / \mathrm{d} t+
\nabla{\bf j}({\bf r}, t)=0$.
Thus, the variational principle Eq.\ (\ref{variation}) with a local function 
$Q({\mathbf r})$ describes excitations by intrinsic local currents. The
time dependence of the parameter $\alpha$ is obviously harmonic,
i.e., $\alpha(t)\propto \cos(\omega_\nu t)$, since the present derivation is
based on linear response theory.

The physical significance of the variational approach now being clear, it
remains to derive the actual equations that determine the displacement
fields $\mathbf{ u(r)}$ and the energies $\hbar \omega$ that are
associated with particular 
excitations. Starting from Eq.\ (\ref{variation2}) and using an explicit
notation, 
\begin{equation}
\frac{\delta m_3[{\mathbf u}[Q({\mathbf r})]]}{\delta Q({\mathbf r}')}-
   (\hbar \omega_1)^2 \frac{\delta
   m_1[{\mathbf u}[Q({\mathbf r})]]}{\delta Q({\mathbf r}')} = 0 =
\int {\mathrm d}^3r'' \left\{
 \frac{\delta m_3[\mathbf{u}({\mathbf r})]}{\delta \mathbf{u}({\mathbf r}'')}-
  (\hbar \omega_1)^2 
  \frac{\delta m_1[\mathbf{u}({\mathbf r})]}{\delta \mathbf{u}({\mathbf r}'')} \right\}  
 \frac{\delta \mathbf{u}({\mathbf r}'')}{\delta Q({\mathbf r}')}
\label{cl1}
\end{equation}
follows by virtue of the chain rule for functional derivatives. Thus,
solutions of 
\begin{equation}
\label{variation3}
 \frac{\delta m_3[{\mathbf u}({\mathbf r})]}{\delta {\mathbf u}({\mathbf r}')}=
  (\hbar \omega_1)^2 
  \frac{\delta m_1[{\mathbf u}({\mathbf r})]}{\delta {\mathbf u}({\mathbf r}')}
\end{equation}
will also be solutions to Eq.\ (\ref{variation2}) and thus Eq.\
(\ref{variation}). $m_1$ is already given as the functional
$m_1[{\mathbf u}]$ by Eq.\ 
(\ref{m1u}), and $m_3[{\mathbf u}]$ is readily obtained by 
inserting the scaled Kohn-Sham orbitals and density from Eq.\ (\ref{scalen})
into the energy functional Eq.\ (\ref{clfunctional}) and calculating the
second derivative with respect to the parameter $\alpha$, Eq.\ (\ref{todft}).
The final equations are then derived in a lengthy but
straightforward calculation from Eq.\ (\ref{variation3}) 
by explicitly performing the variation on $\bf u$. Using the usual
definition
\begin{equation}
\frac{\delta m_3[{\mathbf u}({\mathbf r})]}{\delta {\mathbf u}({\mathbf r}')}=
\frac{\delta m_3[{\mathbf u}]({\mathbf r})}{\delta u_x({\mathbf r}')}
{\mathbf e}_x+
\frac{\delta m_3[{\mathbf u}({\mathbf r})]}{\delta u_y({\mathbf r}')}
{\mathbf e}_y+
\frac{\delta m_3[{\mathbf u}({\mathbf r})]}{\delta u_z({\mathbf r}')}
{\mathbf e}_z,
\end{equation}
where ${\mathbf e}_i$ are the unit vectors in the Cartesian directions, a set
of three coupled, partial differential eigenvalue equations of fourth order
for the Cartesian components $u_j({\mathbf r})$ is obtained:
\begin{equation}
\label{firstfde}
\frac{\delta m_3[{\mathbf u}]}{\delta u_j({\mathbf r})} 
= (\hbar \omega)^2 \,
\frac{\delta m_1[{\mathbf u}]}{\delta u_j({\mathbf r})}, \hspace{0.5cm}j=1,2,3,
\end{equation}
where
\begin{equation}
\frac{\delta m_1[{\mathbf u}]}{\delta u_j({\mathbf r}) }=
\frac{m}{\hbar^2}n({\mathbf r})u_j({\mathbf r}),
\end{equation}
\begin{equation}
\frac{\delta m_3[{\mathbf u}]}{\delta u_j({\mathbf r}) } 
=
\frac{\delta m_3^{\mathrm kin}[{\mathbf u}]}{\delta u_j({\mathbf r}) } 
+
\frac{\delta m_3^{\mathrm KS}[{\mathbf u}]}{\delta u_j({\mathbf r}) } 
+
\frac{\delta m_3^{\mathrm h2}[{\mathbf u}]}{\delta u_j({\mathbf r}) } 
+
\frac{\delta m_3^{\mathrm xc2}[{\mathbf u}]}{\delta u_j({\mathbf r}) },
\end{equation}
and
\begin{eqnarray}
\frac{\delta m_3^{\mathrm kin}[{\mathbf u}]}{\delta u_j({\mathbf r}) }
&=&
-\frac{\hbar^2}{2m}\frac{1}{2}
\sum_{m=1}^{N_{\mathrm e}} \sum_{i=1}^3  \Re {\mathrm e}
\Bigg\{  \Big( \Delta \psi_m \Big)
\bigg[(\partial_ju_i) (\partial_i \psi_m^*) +
(\partial_j \partial_i u_i) \psi_m^* +  u_i(\partial_j \partial_i\psi_m^*)
\bigg]+
\nonumber \\ &&
\bigg[(\partial_j u_i ) (\partial_i \Delta \psi_m) 
+ u_i( \partial_j \partial_i \Delta \psi_m)
\bigg]\psi_m^* -
u_i \bigg[ 
  (\partial_i \psi_m^*) \Big(\partial_j \Delta \psi_m \Big) +
  \Big(\partial_i \Delta \psi_m \Big) 
    (\partial_j \psi_m^*)
\bigg] 
\nonumber \\ && 
  + 2\Bigg[ (\partial_j \psi_m^*)\bigg[ \Delta \Big( 
    \frac{1}{2}(\partial_i u_i)\psi_m + u_i(\partial_i \psi_m)
  \Big)\bigg]-
\bigg[\partial_j \Delta\Big(
\frac{1}{2}(\partial_i u_i)\psi_m + u_i(\partial_i \psi_m)\Big)
\bigg]\psi_m^*\Bigg] \, \Bigg\},
\end{eqnarray}
\begin{equation}
\label{m3ks}
\frac{\delta m_3^{\mathrm KS}[{\mathbf u}]}{\delta u_j({\mathbf r}) }=
\frac{1}{2}\sum_{i=1}^3\Bigg[ n
  \Big( (\partial_j u_i)(\partial_i v_{\mathrm KS})
       -(\partial_i u_i)(\partial_j v_{\mathrm KS})  \Big) +
  u_i \Big( n (\partial_i \partial_j v_{\mathrm KS})
       -(\partial_i n)(\partial_j v_{\mathrm KS})  \Big) \Bigg],
\end{equation}
\begin{equation}
\frac{\delta m_3^{\mathrm h2}[{\mathbf u}]}{\delta u_j({\mathbf r}) }=
n\int\bigg[\sum_{i=1}^3 (\partial_i' u_i({\mathbf r}'))n({\mathbf r}') + 
u_i({\mathbf r}')(\partial_i' n({\mathbf r}'))\bigg]
\frac{r_j-r_j'}{|{\mathbf r}-{\mathbf r}'|^3} \,{\mathrm d}^3r',
\end{equation}
\begin{equation}
\label{lastfde}
\label{vardftm3xc2}
\frac{\delta m_3^{\mathrm xc2}[{\mathbf u}]}{\delta u_j({\mathbf r}) }=
-n\sum_{i=1}^3 \Bigg[ \bigg( \partial_j  ((\partial_i u_i)n + 
u_i (\partial_i n)) \bigg) 
\frac{\partial v_{\mathrm xc}}{\partial n}
+\bigg( (\partial_i u_i) n +u_i(\partial_i n)\bigg)
\bigg(\partial_j \frac{\partial v_{\mathrm xc}}{\partial n}
\bigg)\Bigg],
\end{equation}
where we used the shorthand notation $\partial_1=\partial / \partial x$
etc., and indicated the terms to which derivatives refer by including
them in parenthesis. The usual Kohn-Sham and exchange-correlation
potential are denoted by $v_{\mathrm KS}$ and $v_{\mathrm xc}$, respectively.  

Eqs.\ (\ref{firstfde}) -- (\ref{lastfde}) are our quantum fluid-dynamical
equations. In analogy to the local density approximation (LDA) used for 
$v_{\mathrm xc}$, we term our scheme
the {\it local current approximation (LCA)} to the dynamics, due to the
use of a local function $Q({\mathbf r})$ in the variational principle
(\ref{variation}). It should be noted that the above equations
differ from the equations derived earlier in a semiclassical approximation
\cite{mb89} or by explicit particle-hole averaging \cite{lrpa}.
Due to the fact that our approach is completely based on the Kohn-Sham
density functional and therefore contains the full quantum-mechanical
shell effects in the ground-state density, it is also different from
some fluid-dynamical approaches developed in nuclear physics 
\cite{providencia} (and used in cluster physics \cite{providencia2}) 
which involved either schematic liquid-drop model densities or semiclassical
densities derived from an extended Thomas-Fermi model.

Although Eqs.\ (\ref{firstfde}) -- (\ref{lastfde}) look rather
formidable, they can be solved numerically with reasonable
computational effort, and we have done so for the sodium clusters
${\mathrm Na}_2$ and ${\mathrm Na}_5^+$. The Kohn-Sham equations were
solved basis-set free on a three-dimensional Cartesian real-space grid
using the damped gradient iteration with multigrid relaxation
\cite{diss}.  The ionic coordinates were obtained by minimizing the
total energy using a smooth-core pseudopotential \cite{bigpp}. For
$E_{\mathrm xc}$, we employed the LDA functional of Ref.\
\cite{pw}. The $u_j({\mathbf r})$ were expanded in harmonic oscillator
wavefunctions and we explicitly enforced Eq.\ (\ref{defu}). The
convergence rate of the expansion can be improved by adding a few
polynomial functions to the basis. By multiplying Eqs.\
(\ref{variation3}) and subsequently (\ref{firstfde})--(\ref{lastfde})
from the left with $\mathbf u$ and integrating over all space, a
matrix equation for the expansion coefficients is obtained which can
be solved using library routines. The square roots of the eigenvalues
then give the excitation energies and from the eigenvectors, the
oscillator strengths can be computed.

It should be noted that for systems as small as the ones studied here,
generalized gradient approximations \cite{ggapbe} and their extensions
\cite{mgga} in general are a better approximation to the exchange and
correlation energy than the LDA. However, in the present case LDA is a
good approximation since the valence electrons in sodium clusters are
strongly delocalized.  Furthermore, the bond length underestimation in
LDA which was shown to strongly influence optical properties
\cite{epjd,thermexp} is corrected to a large extent by using the
smooth-core pseudopotential which by construction gives bondlegths
close to the experimetal ones when used with the LDA \cite{bigpp}.

\begin{figure}[htb]
\begin{center}
\includegraphics[width=10cm]{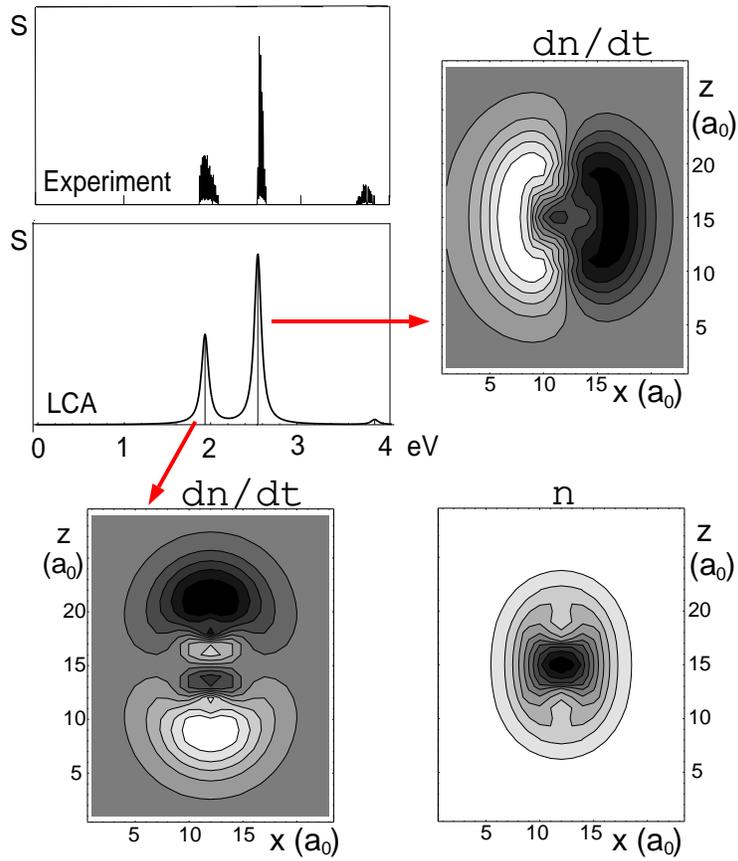}
\end{center}
\caption{From top left to bottom right: Experimental photoabsorption
cross section \protect \cite{na2pabs} and LCA cross section S of
\protect ${\mathrm Na}_2$ in arbitrary units versus eV, contour plots
of the density change associated with the first excitation, the second
excitation, and contour plot of the ground-state valence electron
density. The length unit for the axes of the contour plots is the Bohr
radius $a_0$.}
\label{fig1}
\end{figure}

Fig.\ \ref{fig1} shows the experimental photoabsorption spectrum
\cite{na2pabs} of ${\mathrm Na}_2$ in the upper left picture (adapted
from Ref.\ \cite{chelikowsky}), and below the spectrum obtained in the
just described LCA. We introduced a phenomenological line broadening
in the LCA results to guide the eye. The LCA correctly reproduces the
electronic transitions, despite the fact that only two electrons are
involved. Due to Eq.\ (\ref{seecon}), one can very easily visualize
how the electrons move in a particular excitation by plotting the
corresponding $\nabla \mathbf{j(r)}$, giving a ``snapshot'' picture of
$\mathrm{d} n/ \mathrm{d}t$. For the two main excitations, a
crossection of this quantity along the symmetry axis ($z$ axis) is
shown in the lower left and upper right contourplots, and the
ground-state valence electron density is shown in the lower right for
reference. In the plots of $\mathrm{d} n/ \mathrm{d}t$, shadings
darker than the background grey indicate a density increase, lighter
shadings indicate a decrease. It becomes clear that the lower
excitation corresponds to a density oscillation along the $z$ axis
whereas the higher excitation corresponds to two energetically
degenerate oscillations perpendicular to the symmetry axis. (For the
sake of clarity, we plotted the corresponding oscillator strengths on
top of each other in the photoabsorption spectrum.) This is exactly
what one would have expected intuitively. But the plots reveal that
besides the expected general charge transfer from one end of the
cluster to the other, the presence of the ionic cores hinders the
valence electrons to be shifted freely, creating a density shift of
reverse sign in between the ionic cores.

\begin{figure}[htb]
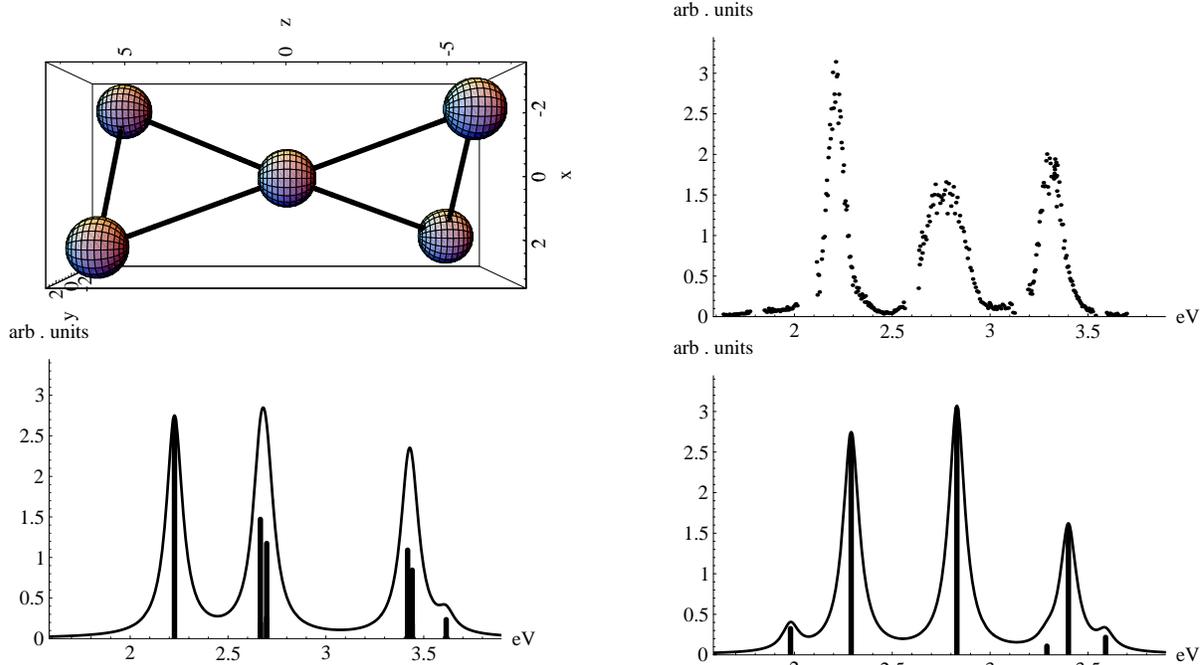

\begin{minipage}{7cm}
\hspace*{0.4cm}
\includegraphics[angle=90,width=7cm]{struct5p.epsi}
\includegraphics[width=7cm]{na5plca.epsi}
\end{minipage}
\hfill
\begin{minipage}{7cm}
\includegraphics[width=7cm]{na5pexp.epsi}
\includegraphics[width=7cm]{na5pkout.epsi}
\end{minipage}
\caption{Upper left: ionic ground-state configuration of
$\mathrm{Na}_5^+$, lower left: corresponding LCA photoabsorption spectrum,
upper right: experimental low-temperature photoabsorption spectrum
\protect \cite{divhaberland}, lower right: Configuration-Interaction
photoabsorption spectrum from Ref.\ \protect \cite{koutecky96}. See text for
discussion.}  
\label{fig2}
\end{figure}

Fig.\ \ref{fig2} shows the ionic ground-state configuration of
$\mathrm{Na}_5^+$ with our labeling of axes in the upper left, the
experimental low-temperature ($\approx$ 100 K) photoabsorption spectrum
\cite{divhaberland} in the upper 
right, the LCA photoabsorption spectrum in the lower left, and the CI spectrum
adapted from Ref.\ \cite{koutecky96} in the lower right. Again, a
phenomenological line broadening was introduced in the presentation of
both the LCA and the CI results.
The LCA spectrum again is in close agreement with the experimentally
observed spectrum, showing three intense transitions. With our choice
of the coordinate system, the lowest excitation corresponds to a density
oscillation in $z$ direction, whereas the two higher excitations
oscillate in both $x$ and $y$ directions. In the interpretation of the
LCA results, it must be kept in mind that
due to our finite grid spacing the numerical accuracy for the
excitation energies is about 0.03 eV, which is absolutely sufficient
in the light of the physical approximations that we are making. But
due to this finite numerical resolution and the fact that we evaluate
each direction of oscillation separately, the $x$ and $y$ components
of the excitations at 2.7 eV and 3.4 eV, which really should be
degenerate for symmetry reasons, appear as extremely close-lying
double lines. However, since the symmetry of the cluster was in no way
an input to our calculation, it is a reassuring test that the LCA, indeed,
fulfills the symmetry requirement within the numerical accuracy. 
Furthermore, it is reassuring to see that with respect to
the relative heights of the peaks the LCA is very close to the CI
results, with differences observed only in the small subpeaks that are
not seen experimentally anyway. And small differences to the CI
calculation are already to be expected simply because of the use
of different pseudopotentials and the resulting small differences in
the ground-state structure.

\section{Conclusion}

In summary, we have derived a set of quantum fluid-dynamical equations
from a general variational principle for the excitations of a many-body
system. The equations describe here the eigenmodes of the system's
(valence) electrons and require only the knowledge of the occupied
ground-state Kohn-Sham orbitals. From these equations, we have computed   
the photoabsorption spectra for small sodium clusters and find
quantitative agreement with the experimentally observed
peak positions. Thus, even low-temperature photoabsorption spectra can
be understood in an intuitive picture of density oscillations, without
knowledge of the true (or any approximate) many-body wavefunction.

\acknowledgments
We are grateful to P.-G. Reinhard for his vivid interest in this work
and for many stimulating discussions. This work was supported by the
Deutsche Forschungsgemeinschaft under grant No.\ Br 733/9 and by an
Emmy-Noether scholarship. S.K.\ is grateful to J.\ Perdew for a warm
welcome at Tulane University.

\begin{references}


\bibitem{hkks}P.\ Hohenberg und W.\ Kohn, Phys.\ Rev.\ {\bf
  136},   B864 (1964); W.\ Kohn und L.\ J.\ Sham, Phys.\ Rev.\ {\bf 140},
  A1133 (1965). 

\bibitem{tddft2}For an overview see, e.g., E.\ K.\ U.\ Gross, J.\ F.\
  Dobson, and M.\ Petersilka,  
  in {\it Density Functional Theory}, edited by R.\ F.\ Nalewajski
  (Topics in Current Chemistry, Vol.\ 181, Springer, Berlin, 1996).

\bibitem{deb}B.\ M.\ Deb and S.\ K.\ Gosh, J.\ Chem.\ Phys.\ {\bf 77}, 342
  (1982). 

\bibitem{vignale}G.\ Vignale, C.\ A.\ Ullrich, and S.\ Conti,
  Phys.\ Rev.\ Lett.\ {\bf 79}, 4878 (1997).

\bibitem{overview}For cluster excitations calculated in DFT, see,
e.g., W.\ Ekardt, Phys.\ Rev.\ B {\bf 31}, 6360 (1985); M.\ Madjet, C.\
  Guet, and W.\ R.\ Johnson, Phys.\ Rev.\ A {\bf 51}, 1327 (1995); U.\
  Saalmann and R.\ Schmidt, Z.\ Phys.\ D {\bf 38}, 153 (1996); A.\ Rubio,
  J.\ A.\ Alonso, X.\ Blase, L.\ C.\ Balb\'{a}s, and S.\ G.\ Louie,
  Phys.\ Rev.\ Lett.\ {\bf 77}, 247 (1996); K.\ Yabana and G.\ 
  F.\ Bertsch, Phys.\ Rev.\ B {\bf 54}, 4484 (1996); 
  A.\ Pohl, P.-G.\ Reinhard, E.\ Suraud, Phys.\ Rev.\ Lett.\ {\bf 84},
  5090 (2000).   

\bibitem{chelikowsky}
  I. Vasiliev, S. \"O\u{g}\"ut, and J. R. Chelikowsky, Phys.
  Rev. Lett. {\bf 82}, 1919 (1999). 

\bibitem{hannupalad}M. Moseler, H. Hakkinen, R.N. Barnett, and U.
  Landman, to appear in Phys. Rev. Lett. 2001; lanl preprint physics/0101069.

\bibitem{akola00}J. Akola, A. Rytk\"onen, H. H\"akkinen, and
  M. Manninen, Eur. Phys. J. D {\bf 8}, 93 (2000).

\bibitem{bigpp}S. K\"ummel, M. Brack, and P.-G. Reinhard,
  Phys. Rev. B {\bf 62}, 7602 (2000).

\bibitem{divhaberland}C. Ellert, M. Schmidt, C. Schmitt, T. Reiners,
  and H. Haberland, Phys. Rev. Lett. 75, 1731 (1995); 
  M. Schmidt, C. Ellert, W. Kronm\"uller, and H. Haberland,
  Phys. Rev. B {\bf 59}, 10970 (1999). 

\bibitem{koutecky96}V. Bona\v{c}ic-Kouteck\'{y}, J. Pittner, C. Fuchs,
  P. Fantucci, M. F. Guest, and J. Kouteck\'{y}, J. Chem.  
  Phys. {\bf 104}, 1427 (1996).

\bibitem{rowe}We particularly like the presentation of this technique given in
  D. J. Rowe, {\it Nuclear collective motion} (Methuen and Co., London,
  1970). 

\bibitem{lrpa}P.-G.\ Reinhard, M.\ Brack and O.\ Genzken, Phys.\ 
  Rev.\ A {\bf 41}, 5568 (1990).

\bibitem{mb89}M. Brack, Phys. Rev. B {\bf 39}, 3533 (1989).

\bibitem{bohigas}O. Bohigas, A. M. Lane, and J. Martorell,
  Phys.  Rep. {\bf 51}, 267 (1979); and references therein.

\bibitem{providencia}E.\ R.\ Marshalek and J.\ da Provid\^{e}ncia, Phys.\
  Rev.\ C {\bf 7}, 2281 (1973); J.\  da Provid\^{e}ncia and G. Holzwarth,
  Nucl.\ Phys.\ A {\bf 439}, 477 (1985); E. Lipparini and
  S. Stringari, Phys.\ Rep.\ {\bf 175}, 103 (1989);
  P. Gleissl, M. Brack, J. Meyer, and P. Quentin, Ann.\ Phys.\ (N.Y.)
  {\bf 197}, 205 (1990). 

\bibitem{providencia2}J.\ da Provid\^{e}ncia, Jr.\, and
  R.\ de Haro, Jr., Phys.\ Rev.\ B {\bf 49}, 2086 (1994).

\bibitem{diss}V.\ Blum, G.\ Lauritsch, J.\ A.\ Maruhn, and  
  P.-G.\ Reinhard, J.\ of Comp.\ Phys.\ {\bf 100}, 364 (1992); 
  S.\ K\"ummel, {\it Structural and  Optical Properties of Sodium
Clusters studied in Density Functional Theory}, (Logos Verlag, Berlin, 2000).

\bibitem{pw}J.\ P.\ Perdew and Y.\ Wang, Phys.\ Rev.\ B {\bf 45},
  13244 (1992).

\bibitem{ggapbe}J.\ P.\ Perdew, K.\ Burke, and M.\ Ernzerhof, Phys.\
  Rev.\ Lett.\ {\bf 77}, 3865 (1996).

\bibitem{mgga}J.\ P.\ Perdew, S.\ Kurth, A.\ Zupan, and P. Blaha,
Phys.\ Rev.\ Lett.\ {\bf 82}, 2544 (1999).

\bibitem{epjd}S.\ K\"ummel, T.\ Berkus, P.-G.\ Reinhard, and M.\
  Brack, Eur.\ Phys.\ J.\ D {\bf 11}, 239 (2000).

\bibitem{thermexp}  S.\ K\"ummel, J.\ Akola, and M.\ Manninen, Phys.\ Rev.\
  Lett.\  {\bf 84}, 3827 (2000).

\bibitem{na2pabs}W.\ R.\ Fredrickson and W.\ W.\ Watson, Phys.\ Rev.\
  {\bf 30}, 429 (1927).


\end{references}
\end{document}